\documentclass[12pt]{article}
\pdfoutput=1
\usepackage{amssymb}
\usepackage{amsmath}
\usepackage{graphicx}
\usepackage{hyperref}
\usepackage{color}

\newcommand{\eq}{\begin{equation}}
\newcommand{\eqx}{\end{equation}}
\newcommand{\eqs}{\begin{equation*}}
\newcommand{\eqsx}{\end{equation*}}
\newcommand{\eqn}{\begin{eqnarray}}
\newcommand{\eqnx}{\end{eqnarray}}
\newcommand{\eqns}{\begin{eqnarray*}}
\newcommand{\eqnsx}{\end{eqnarray*}}

\newcommand{\f}[2]{\frac{#1}{#2}}

\newcommand{\Sg}{\Sigma}
\newcommand{\dl}{\delta}

\newcommand{\al}{\alpha}

\newcommand{\qqqq}{\quad\quad\quad\quad}
\newcommand{\qq}{\quad\quad}

\newcommand{\nn}{{\cal N}}

\title{Towards holography for quantum mechanics}

\author{Romuald A. Janik\thanks{e-mail: {\tt romuald.janik@gmail.com}} \\ \\ 
\small 
Institute of Physics\\\small
Jagiellonian University\\\small
ul. {\L}ojasiewicza 11, 30-348 Krak{\'o}w, Poland}

\date{}

\begin{document}

\maketitle

\begin{abstract}
We derive a holographic description of the simplest quantum mechanical system,
a 1d free particle. The dual formulation uses a couple of two-dimensional topological abelian BF theories with appropriate
boundary conditions, interactions and constraints.
The aim of this construction is not to use holography as a tool for quantum mechanics but rather
to find the simplest possible setup in order to explore holography.
\end{abstract}

\vfill

\pagebreak

\section{Introduction}

The AdS/CFT correspondence \cite{ADSCFT} is one of the most fascinating theoretical breakthroughs
in recent years. In its original form it postulated the duality between two
completely different theories -- the supersymmetric gauge theory $\nn=4$ Super-Yang-Mills theory
in 4 dimensions and superstring theory on a $AdS_5 \times S^5$ background. Since then it has been
extended in numerous directions. Apart from very important practical applications as a tool
for learning about the nonperturbative dynamics of gauge theory it is particularly
fascinating theoretically as it proposes the equivalence of a nongravitational theory
(the gauge theory ``on the boundary'') and quantum theory incorporating gravity.
This is a very explicit realization of the holographic principle~\cite{HOLOGRAPHY}.
However the very reason for which the AdS/CFT correspondence is so useful a tool for studying nonperturbative gauge
theory physics makes it difficult to understand its origin microscopically
from the gauge theory point of view. Indeed both sides of the duality become 
simple in opposite limits. In particular 
we do not know how to deal with string theory in the quantum gravity regime 
corresponding to small coupling \emph{and} finite number of colors on the gauge theory side.
From the point of view of understanding holography the optimal setup would be
to have relatively simple and tractable quantum theories on both sides of the duality.

Some particularly intriguing generalizations of holography involved three dimensional free $O(N)$ vector model
which was proposed to be dual to four dimensional Vasiliev gravity \cite{ON, 4DHS}.
A lot of progress was made in the understanding of dual dynamics from the boundary
theory point of view (see e.g. \cite{JEVICKI, LEIGH}), however the gravitational
side is basically understood only at the classical and semi-classical level as
Vasiliev gravity \cite{VASILIEV} has not been quantized so far.

Reducing the number of dimensions, a class of two dimensional CFT's was proposed
to be dual to three dimensional Vasiliev gravity coupled to a scalar field \cite{GABGOP}.
In this case there is an explicit action for the Vasiliev theory which is a difference of
two Chern-Simons theories, however the total action incorporating interactions with the scalar field is unknown and it is very
difficult to study the duality in the finite $k$, finite $N$ case\footnote{In this case $k$
and $N$ are the parameters of the 2D coset CFT's.}.

Most recently the Sachdev-Ye-Kitaev (SYK) model \cite{SYK} (see also \cite{MALDACENASTANFORD} and subsequent developments) became intensively studied as 
it is a quantum mechanical system which exhibits properties characteristic of 
a dual holographic classical gravity description in terms of black holes.

In another line of investigation, it was realized that entanglement is crucially connected
with holography.
Surprising parallels were uncovered between 
the description of ground state wave functions using MERA (Multi-scale Entanglement Renormalization Ansatz) \cite{SWINGLE,TAKAYANAGI}
and the Ryu-Takayanagi holographic prescription for computing entanglement entropy \cite{RYUTAKAYANAGI}.
More recently various models for holography were proposed incorporating various tensor network constructions
in particular the HaPPY proposal taking into account spatial error correcting features of the holographic dictionary \cite{HAPPY}.
Other recent advances include a path integral optimization framework \cite{OPTIM} and
the random tensor networks \cite{RANDOMTENSOR}.

One generic feature of the approach to understand holography in terms of tensor networks is that
these constructions are in a sense very kinematical. E.g. the HaPPY proposal provides a mapping of a boundary Hilbert space to a bulk Hilbert space
which is quite agnostic about the dynamics (Hamiltonian/action etc.) of the boundary theory. 
If this intuition is true, it suggests that a holographic description should be in principle applicable to almost any system\footnote{By
a holographic description we mean throughout this paper a generic higher dimensional dual description which may be
very quantum and far from a description in terms of classical gravity. So we use the term in a much wider
sense than e.g. in \cite{POLCHINSKI}.}.

In this short note we would like to investigate whether one can formulate a holographic dual model for the arguably simplest
possible quantum system -- a free particle in 1 dimension. If successful, this could be a starting point of studying
more complicated setups with more degrees of freedom, interactions etc. in a context which is very much under control.

The plan of the paper is as follows. First we review some very basic requirements for a holographic description
of a given theory and for identifying a gravitational subsector of the holographic bulk theory. 
Then we proceed to implement this program for the quantum mechanical free particle.
We close the paper with a summary and conclusions.

\section{The main features of a holographic description}
\label{s.features}

In this section we will summarize what we would expect from a holographic description of some theory.
Suppose that the field theory in question is defined in $d$ spacetime dimensions on some fixed
nondynamical geometry $\Sg$. 

\bigskip

{\bf I.} The dual holographic theory should be defined on a higher dimensional
manifold $M$, having $\Sigma$ as  a boundary. At the very least we should be able to match partition
functions for the two theories
\eq
Z_{boundary} = Z_{bulk}
\eqx

{\bf II.} The above requirement is not really enough as we should expect to be able to link all correlation functions in the boundary theory
to the bulk theory through the GKP formula \cite{GKP, WITTEN}. Observables/operators in the boundary theory should be
associated to fields in the bulk theory. Moreover the corresponding sources in the generating function
of correlators in the boundary theory should be linked to the boundary values of the associated bulk fields\footnote{For simplicity we ignore potential $z^\Delta$ 
factors and assume that they have been incorporated in a redefinition of the bulk fields.}
namely
\eq
\label{e.genfunc}
\int D\phi\; e^{iS_{bndry}(\phi)+i\int_\Sg j(x^\mu) O(x^\mu) d^dx} = Z_{bulk}\left(\Phi_O(z,x^\mu) \underset{z\to 0}{\longrightarrow} j(x^\mu)\right)
\eqx
Ultimately the boundary degrees of freedom would have been integrated out and the remaining vestiges of the boundary theory
would be just the sources i.e. boundary values of the bulk fields.

{\bf III.} Finally we would like to interpret a part of the bulk theory as a gravitational theory.
In all holographic constructions so far, 
the bulk metric is the field associated to the energy momentum tensor of the boundary theory.
In other words its boundary values should be linked in some way\footnote{We are purposefully quite vague about the details here.
In standard AdS/CFT the dictionary is clearest in the Fefferman-Graham coordinates \cite{SKENDERIS}.
We do not want to impose \emph{a-priori} any specific prescription in the general case.} to the \emph{nondynamical} metric of
the boundary theory. 
Of course, as in the case of higher spin gravity the whole picture may be more complex
with other massless higher spin fields making the geometric interpretation ambiguous,
but still in this way we may identify a natural gravitational subsector of the bulk theory.

\section{A holographic description of a quantum mechanical free particle}

The goal of this note is to try to satisfy the above requirements for one of the simplest systems possible,
the quantum mechanical free particle in one dimension. 
\emph{A-priori} it is not at all clear if such a description exists for such a simple system.
If it does exist, it may well be that the outcome is too trivial and restricted, but we hope that even such failure
may be instructive and interesting as it may indicate a sharpening of the requirements for holography
with respect to the ones outlined in the preceding section. From another perspective it may be
a starting point for constructing holography for more nontrivial quantum mechanical systems.

This system can be understood as a QFT with no spatial
dimension with the action
\eq
S = \int dt\; \f{1}{2} \dot{q}^2
\eqx
Since this system as it stands does not have any coupling
or large $N$ parameter we expect the dual bulk theory to be necessarily quantum. This is in fact one of the key 
motivations of this study. We will now build up the bulk theory in steps in order to satisfy the three
requirements described in section~\ref{s.features}.

\subsection*{The partition function}

Let us consider a two-dimensional abelian BF theory defined on the half plane 
\eq
M=\{(t,z): z \geq 0\}
\eqx
The action is given by 
\eq
S_{BF}= \int_M B dA = \int B(\partial_t A_z - \partial_z A_t) dt dz
\eqx
We would like to impose the following boundary conditions:
\eq
\label{e.Abc}
B=-A_t\; |_{z=0} \qqqq A_t = 0\; |_{z \to \infty}  
\eqx
In order for these boundary conditions to be consistent with the variational principle we have to add to the action a boundary term
\eq
S_{bulk}^I = S_{BF} +\f{1}{2} \int_{\{z=0\}} B^2 dt
\eqx
The variation of the action is now
\eq
\dl S_{bulk}^I = (EOM's) + \int_{\{z=0\}} B \dl A_t dt +  \int_{\{z=0\}} B \dl B dt
\eqx
which vanishes due to the boundary condition $\dl A_t+\dl B = 0 |_{z=0}$.
The superscript on $S_{bulk}^I$ indicates that this will not be the full final bulk action
but will be still modified in the following sections.

Let us now evaluate the bulk action $S_{bulk}^I$.
The Lagrange multiplier field $B$ imposes the constraint that $A$ is a flat connection, hence we may set
\eq
A_z = -\partial_z \Phi \qqqq A_t = -\partial_t \Phi
\eqx
The bulk part of the action $S_{bulk}^I$ on the constraint surfaces vanishes and we are left with just the boundary term 
given through the $B$ field, which in turn due to our boundary conditions
can be expressed in terms of the temporal derivatives of the boundary values of $\Phi(t,z)$ field
\eq
\dot{q}(t) = \lim_{z \to 0} \partial_t \Phi(t,z)
\eqx
We thus reproduce the quantum mechanical free particle action\footnote{Similar computations as in this subsection have been done independently with different
motivations in \cite{BFQM1,BFQM2} in the case of nonabelian BF theories.}.
\eq
\int dt\; \f{1}{2} \dot{q}^2
\eqx
The above simple derivation is a two-dimensional analog of the three-dimensional link of Chern-Simons and 2d WZW \cite{CS1},
in the variant where the boundary conditions are $A_+ = \bar{A}_-=0$ (see e.g. \cite{CS2}).

\subsection*{Source for $q(t)$}

Let us now generalize the construction by adding a generic time dependent source for the particle position $q(t)$.
We thus have to reproduce an additional term in the boundary action
\eq
\int dt\; \f{1}{2} \dot{q}^2 + \int dt\; j(t)q(t)
\eqx
In terms of the BF theory gauge field, the particle position $q(t)$ can be understood essentially as a Wilson line
extending from the boundary to the interior at $z=\infty$ as we have
\eq
\int_{z=0}^\infty A_z\, dz = -\int_{z=0}^\infty \partial_z \Phi(t,z) = \Phi(t,0) - \Phi(t,\infty)
\eqx
Now due to the boundary condition at infinity $A_t = 0\; |_{z \to \infty}$, $\Phi(t,\infty)$ is a constant
and hence without loss of generality can be set to zero. Therefore we can make an identification
\eq
q(t) = \int_L A
\eqx
where the line $L$ is attached to the boundary at time $t$ and goes to infinity in the bulk.
Now we would like to rewrite the integral
\eq
\label{e.source}
\int dt\; j(t)q(t)
\eqx
as a two dimensional integral in terms of natural bulk quantities. We will also need a bulk field
associated to the boundary source $j(t)$.

To this end, we will introduce another two-dimensional abelian BF theory which we will denote by
\eq
\label{e.calpha}
\int C\, d\al
\eqx
In order to write the coupling (\ref{e.source}) we will introduce yet another ingredient: a globally defined
1-form in the bulk which we will denote by $dt$ (for the moment this can be understood as a gradient of the $t$ coordinate). 
\emph{A-priori} the existence of such 1-form in the context
of nonrelativistic quantum mechanics is quite natural in view of Galilean symmetry.
We will, however, return to this point in the following section. For the moment we will treat the 1-form $dt$ as
fixed and given externally as a gradient of the global bulk $t$ coordinate. 

We will now enlarge the bulk action to
\eq
S_{bulk}^{II} = \int_M B\, dA + C\,d\al+ \al \wedge A + D \, \al \wedge dt + \f{1}{2} \int_{\partial M} B^2 dt
\eqx
Integrating over the Lagrange multiplier $D$ restricts the general form of the $\al$ 1-form:
\eq
\al = j(t,z) dt
\eqx
Subsequently integrating over $C$ ensures that $j(t,z)$ is only a function of $t$:
\eq
\al = j(t) dt
\eqx
Now we may evaluate the bulk interaction term between the gauge fields of the two BF theories:
\eq
\int_M \al \wedge A = \int_M j(t) dt \wedge (A_t dt+ A_z dz) = \int j(t) \int_0^\infty A_z dz dt = \int j(t) q(t) dt
\eqx
obtaining exactly the boundary source term for $q(t)$. 

In principle we should now perform the path integral over $A$ leaving an effective bulk action depending on
the scalar fields $B$, $C$, $D$ and gauge field $\al$. We will not attempt to do this in this work but rather
we will return to the 1-form $dt$.

\subsection*{Covariantizing $dt$ and the ``gravity'' subsector}

Since the quantum mechanical path integral is essentially just a QFT on a 1-dimensional worldline, one can
introduce a fixed 1-dimensional metric $g_{tt}(t)$ and write the action as
\eq
\f{1}{2} \int \sqrt{g}\, g^{tt} (\partial_t q)^2 = \f{1}{2} \int \f{1}{e} \dot{q}^2
\eqx 
where we introduced the standard einbein notation, and $e=e(t)$ is a \emph{fixed} given function of time.

We would now like to complete the program sketched in section \ref{s.features} and introduce a bulk
field which would go over to the einbein on the boundary. At the same time we will get rid of the rather
artificial looking external 1-form $dt$ which was necessary to write the boundary source term in terms of
bulk fields. Since $dt$ understood as the gradient of the global bulk time coordinate is necessarily
a closed 1-form, it is extremely suggestive to consider it as a gauge field of a third abelian BF theory
which we will denote by
\eq
\label{e.eeta}
\int E\, d\eta
\eqx   
As the boundary condition at the physical boundary $z=0$ we will fix the temporal component of $\eta$
\eq
\eta = \eta_t dt + \eta_z dz
\eqx
to a fixed value which we will identify shortly with the eibein $e(t)$.
More precisely we fix the pullback of $\eta$ to the boundary $\partial M$ to be equal to $e(t)dt$.
Thus in the case of (\ref{e.eeta}) (as well as for (\ref{e.calpha})) we do not need to add any boundary terms to the action
as was the case for the original $\int B\, dA$ theory.
We will also modify the boundary conditions (\ref{e.Abc}) at $z=0$ to
\eq
\label{e.bcnew}
A_t + \eta_t B = 0 |_{z=0}
\eqx
Accordingly we need to modify the additional boundary term
\eq
\f{1}{2} \int_{\{z=0\}} B^2 dt \longrightarrow \f{1}{2} \int_{\partial M} B^2\, \eta
\eqx
The cancellation of the boundary terms in the variational principle goes through since due to our boundary conditions $\dl \eta_t =0|_{z=0}$.
The resulting boundary action can be seen to be
\eq
\f{1}{2} \int_{\partial M} B^2\, \eta = \f{1}{2} \int \f{1}{\eta_t} A_t^2 dt = \f{1}{2} \int \f{1}{\eta_t} \dot{q}^2
\eqx
where we used (\ref{e.bcnew}). It is now clear that we have to identify the boundary value of $\eta_t$ with the einbein $e(t)$
as announced earlier. From the considerations of section \ref{s.features} we are led to identify the $E$, $\eta$ subsector
as a part of the ``gravitational'' sector of the bulk theory. Note that although this is a two dimensional BF theory
it is distinct from Jackiw-Teitelboim 2D gravity which is a nonabelian BF theory \cite{JACKIW}.

Let us now put together all ingredients introduced so far. Our final bulk action takes the form
\eq
\label{e.sbulkfin}
S_{bulk}^{III} = \int_M B\, dA + C\,d\al+ E\, d\eta + \al \wedge A + D \, \al \wedge \eta + \f{1}{2} \int_{\partial M} B^2 \eta
\eqx
with the boundary conditions at $z=0$
\eq
A_t + \eta_t B = 0 |_{z=0} \qq \al_t =j(t) |_{z=0} \qq \eta_t = e(t) |_{z=0}
\eqx
Let us make some comments on the above expression. Increasing the number of degrees of freedom
will increase the number of components of all fields except $\eta$ and $E$.
Adding interactions (on the quantum mechanical side) is rather nontrivial. One can either integrate over the source
or introduce separate sources for the monomials $q(t)^n$. Doing that seems to require a significant extension
to the formalism. Ultimately we would also like to integrate out $A$ and possibly $B$. 
We leave these issues for future investigation. 

\section{Conclusions}

The motivation for the construction presented in this note is the intuition arising from tensor network
interpretations of holography that a holographic description should exist for almost any system.
Hence it should be possible to find a holographic formulation of the most extreme simple system
that one could think of -- a one dimensional quantum mechanical free particle.
As we would like to have an explicit dual theory described by some concrete bulk action,
we did not take the approach through tensor network constructions but rather we worked directly
in the continuum with two dimensional topological BF theories having the Chern-Simons/WZW relation as a guiding principle.
The expected features of a holographic dual imposes, however, further requirements on the bulk theory going beyond the equality of
partition functions. In particular we should have additional
matter fields in the bulk theory which are associated to the operators of the boundary theory and which reduce to
the corresponding sources at the boundary. In this work we carried out the construction for the source for the particle position~$q(t)$.
We also identified a subsector of the bulk theory which reduces to the einbein on the boundary
and thus behaves like a ``gravitational'' sector of the bulk theory.

A characteristic feature of the simple quantum mechanical model considered here is the absence of
a large $N$ parameter. More precisely, one can consider this model to have $N=1$, with a straightforward
generalization to $N$ components. In the conventional examples of the AdS/CFT correspondence,
finite $N$ corresponds to a quantum bulk model (in these cases quantum gravity+other matter fields),
which was also a motivation for the present construction, where we treat the bulk theory on the quantum level
as we use the full path integral formalism. Indeed the role of a large $N$ limit in a generalized version
of the model (possibly with a singlet constraint) within a similar construction is a very interesting
problem which we plan to address in the future.
 
One qualitative feature of holography which is not explicitly captured by the present construction is the
interpretation of the holographic direction as an RG flow. In the present paper, on the other hand, the starting point of the construction
was a minimal implementation of the bulk formula for the generating function of correlators (\ref{e.genfunc}),
which does not lead to a direct RG interpretation (which in any case is not evident as the quantum mechanical 
system lives on a worldline and thus has no spatial dimension). We suspect that to address this issue
one would have to integrate out the $A$ and $B$ fields and analyze the resulting theory of just the bulk fields
associated with sources of $q(t)$ and the einbein $e(t)$. Possibly for a local geometric interpretation
one would have to combine this procedure with the large $N$ limit discussed above. 
This goes beyond the scope of the present paper but is definitely another important problem for future research.

There are also many other possible directions for further investigation, foremost of which is going to
nontrivial quantum mechanical systems. It is not completely clear whether to consider in addition sources 
for monomials of $q(t)$ and to what extent the construction of the source sector performed
here is unique or optimal. On a more mundane level it would be interesting to analyze
the bulk theory in more detail and check to what extent our experience with holography
in higher number of dimensions carries through here. 
We hope that the setup presented in this paper would be a good framework to address such questions.

\bigskip

\noindent{\bf Acknowledgements.}\\ 
This work was supported by NCN grant 2012/06/A/ST2/00396.

\end{document}